\documentclass[epj]{svjour}
\usepackage{graphicx}
\usepackage{amsmath}
\usepackage{amssymb}

\begin{document}

\title{Structure and Fragmentation in Colloidal 
Artificial Molecules and Nuclei} 
\titlerunning{Structure and Fragmentation...}
\author{C.J. Olson Reichhardt, C. Reichhardt \and A.R. Bishop}
\institute{ 
Theoretical Division, 
Los Alamos National Laboratory, Los Alamos, New Mexico 87545
}

\abstract{
Motivated by recent experiments on colloidal systems with
competing attractive and repulsive interactions, 
we simulate a two-dimensional system of colloids
with competing interactions that can undergo fragmentation. 
In the absence of any
other confining potential, the colloids can form stable clusters
depending on the strength of the short range attractive term.
By suddenly changing the strength of one of the interaction
terms we 
find a rich variety of fragmentation behavior which is affected
by the existence of ``magic'' cluster numbers.
Such soft matter systems can 
be used to construct artificial nuclei. 
\PACS{{82.70.Dd}{Colloids} \and
  {45.70.Qj}{Pattern formation} \and
  {05.70.Ln}{Nonequilibrium and irreversible thermodynamics}
}}
\maketitle

The issue of pattern stability has attracted considerable interest in
a wide range of different areas.  A commonly encountered mechanism
for pattern formation is the competition between short-range
attraction and long-range repulsion, which occurs in many systems
including magnetic films, 
Langmuir monolayers, diblock copolymers \cite{Seul95},
water-oil mixtures \cite{Gelbart96}, and
two-dimensional electron systems \cite{2Delectron}.
Similar types of pattern formation are expected to occur in dense nuclear 
matter, where they have been termed ``pasta'' phases.
These phases may occur under conditions of stellar collapse \cite{Ravenhall83}
or in the crust of a neutron star \cite{Watanabe020405}.
Models of classically interacting particles with competing attractive and
repulsive interactions have been used to study a variety of topics in
nuclear matter.
For example, if two nuclei collide, they temporarily form a high density
state which then expands rapidly and can fracture in a complex manner
\cite{Gross93}.  
Nuclear fragmentation processes in both two-dimensional \cite{Strachan97}
and three-dimensional \cite{Lenk86} geometries  
cause the nucleus to be decomposed into clusters of all sizes
through mechanisms ranging from evaporation
to multifragmentation. 

Models of small numbers of interacting particles, outside the thermodynamic
limit, can be useful in understanding the structure and stability of
naturally occurring clusters such as nuclei.
Repulsively interacting particles can be forced to form clusters by
confining the particles in a trap potential.  A wide variety of systems exhibit
this type of behavior, including electrons in quantum dots \cite{Reed} or
on the surface of liquid helium \cite{Leiderer87}, vortices in
superfluids \cite{Kondo92}, colloidal particles in circular traps
\cite{BubeckWei}, confined ferromagnetic particles \cite{Davidov02},
and charged dust particles in plasma traps \cite{Juan98}.
The resulting structures have
been studied extensively for a range of interaction types and trap
types \cite{Traps}, and the
details are sensitive to the nature of the trapping potential. 

There has recently been growing activity
in studying colloidal systems with
competing long range repulsive and short attractive interactions, where
various types of clusters are observed to form
\cite{Kegel,Stradner,Mossa,Imperio,Kegel2}.
For instance, clusters and ``artificial atoms'' 
were created in an experimental system of 
rotating disks with no confining potential \cite{Stone00Rotating}.
Other current experiments use combinations of entropic
attractive forces with additional repulsive interactions 
to form clusters and study their decay \cite{Grier}, and long-range
repulsion between colloids has been achieved in nonpolar solvents 
\cite{DufresnePrivate} and water-oil emulsions \cite{ChaikinAPS}. 
Analogies have been made between colloidal clusters and nuclear matter,
leading to proposals for mimicking nuclear matter phenomena with a 
colloidal system \cite{Kegel2}. 
Although there have been several colloidal studies of equilibrium
cluster shapes that have identified magic numbers, to our knowledge
the fragmentation of colloidal clusters has not been 
considered previously.

Here, we form clusters out of colloids that have a competing short-range
attraction and a long-range Coulomb repulsion.  
We identify the stability line as a function of the strength of the
short-range attraction for clusters of different sizes.  When the
attraction falls above this line, clusters exist in the absence of a
confining potential.  We then study different mechanisms for destabilizing
the clusters, including thermal excitation and a sudden quench of the
attractive term to a lower value.  The latter process is inspired by
the sudden expansion of an energized nucleus.  
The microscopic details of the decays can be accessed directly through
our simulations.  We find numerous
decay modes depending on the cluster size and the quench depth.
This system can serve as a model for understanding the complex interactions
between charged fragments in a cluster that is breaking apart.

\begin{figure}
\includegraphics[width=0.5\textwidth]{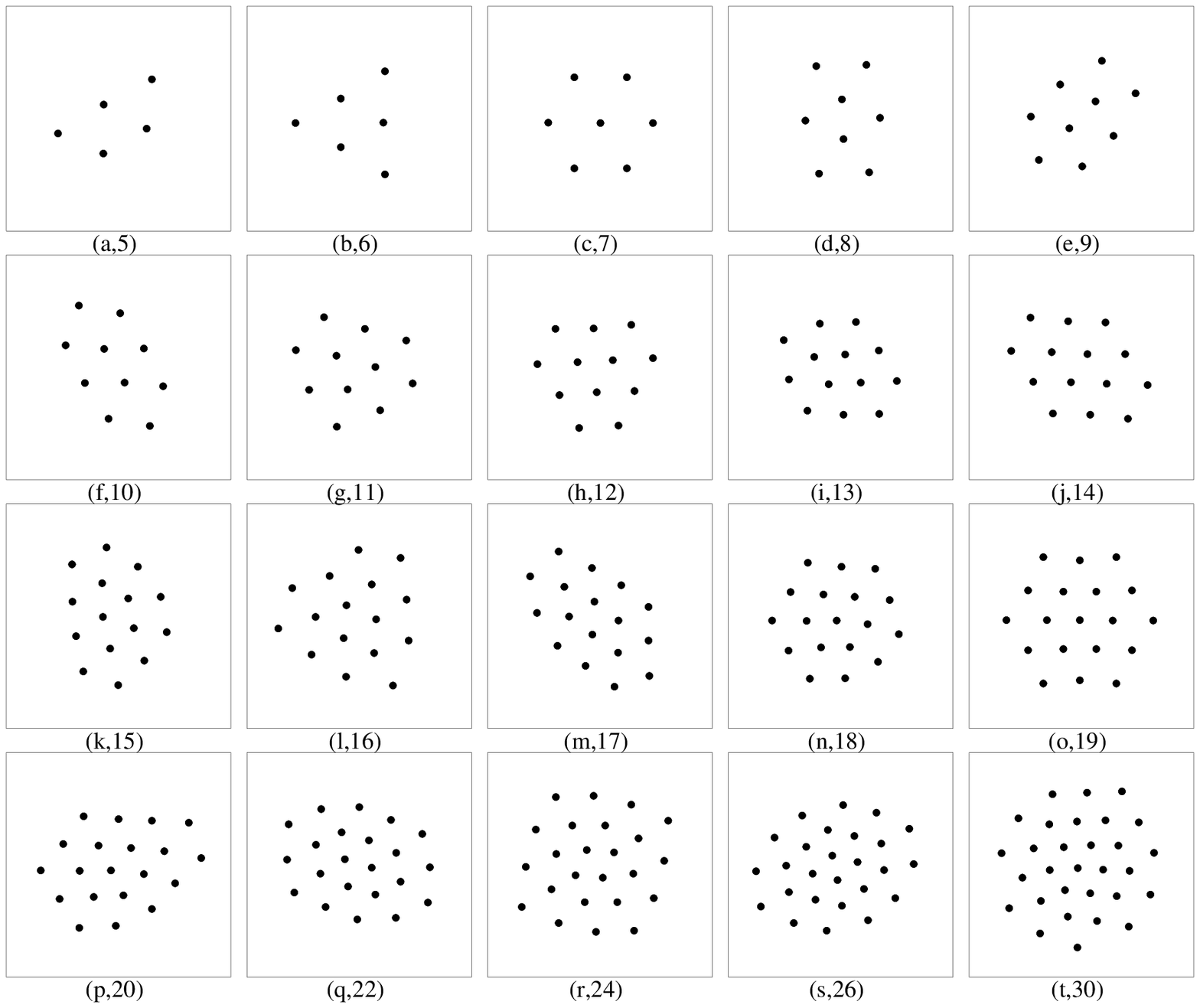}
\caption{Structure of stable clusters for N=(a) 5, (b) 6, (c) 7, (d) 8, (e) 9, 
(f) 10, (g) 11, (h) 12,
(i) 13, (j) 14, (k) 15, (l) 16, (m) 17, (n) 18, (o) 19, (p) 20, (q) 22, 
(r) 24, (s) 26, (t) 30.
}
\label{fig:image}
\end{figure}

We simulate $N$ interacting colloids in a system with open boundary
conditions.  No confining force is applied to the colloids, which
either remain in a cluster due to the colloid-colloid interactions
when placed in a stable configuration, or 
scatter under Coulomb repulsion when placed in an unstable configuration.
The overdamped equation of motion for colloid $i$ is
$\eta d{\bf r}_i/dt = {\bf f}_{i} 
= \sum_{j\ne i}^N{\bf f}_{ij} + {\bf f}^{T},$     
where $\eta$ is a phenomenological damping term.   
The  colloid-colloid interaction 
${\bf f}_{ij}=-\nabla U(r_{ij}){\bf \hat r}_{ij}$ 
for colloids separated by ${\bf r}_{ij}={\bf r}_i-{\bf r}_j$
consists of a long-range 
Coulomb repulsion and a short-range exponential attraction: 
\begin{equation}
U(r) = 1/r - B\exp(-\kappa r).
\end{equation}
At small and large $r$ the repulsive 
Coulomb term dominates. The attractive interaction 
can be varied using the inverse screening length 
$\kappa$ and the parameter $B$.  
Here we fix $\kappa=1$.
As we will demonstrate, clusters are
stabilized above a critical value $B_c$ which is a function of $N$.
The temperature is modeled as Langevin random kicks with the properties
$\langle f^{T}(t)\rangle = 0$ and $\langle f^{T}(t)f^{T}(t^{\prime})\rangle = 
2\eta k_{B}T\delta(t-t^{\prime})$.  To initialize the system, we set
$B \ge B_c$ and allow the colloid positions to relax into a stationary
state in the absence of temperature.  

Due to the attractive component of the colloid-colloid interaction,
stable clusters of colloids form even in the absence of a confining
potential.  We illustrate the stable clusters obtained for 
representative values of $N$ ranging from $N=5$ to $N=30$
in Fig.~\ref{fig:image}.  In general, the colloids tend to form a
small portion of a triangular lattice whenever possible, 
not unlike the hypothetical triangular structure assumed in the nuclear 
lattice model \cite{Bauer85}.
 Since the
underlying substrate is flat, geometrically necessary dislocations do
not arise.  Certain configurations which are the most lattice-like are
also the most stable.  These ``magic'' clusters include $N=7$, 12, and 19.  
Note that the magic cluster $N=12$ forms a nonhexagonal segment of
a triangular lattice, so that it is highly stable even though it does
not meet the proposed magic cluster criterion of
$N=1+3p(p-1)$ for integer $p$ \cite{Jean01}.
The strong tendency toward triangular ordering
is in contrast to systems of confined repulsively interacting particles,
which tend to form ringlike structures instead.  
The smaller clusters shown in Fig.~\ref{fig:image} resemble structures
obtained in the cluster phase of either a periodic system \cite{physicaD} 
or one confined to a trap \cite{nelissen}.  In these systems, interactions
with neighboring clusters alter the shapes of larger clusters and can
distort the triangular ordering.  Thus, the configurations we observe here
for larger unconfined clumps differ from those found in confined or
periodic systems.

\begin{figure}
\includegraphics[width=0.5\textwidth]{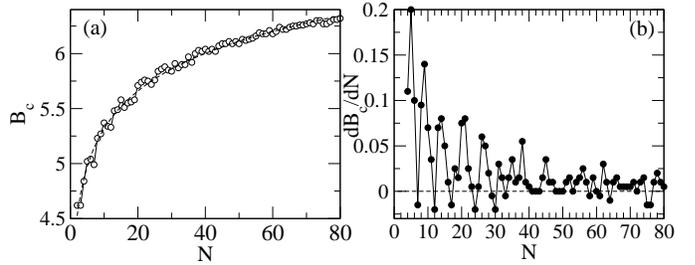}
\caption{(a) Dots: 
Critical value $B_c$ for stable cluster formation versus
cluster size N.  Dotted line: Fit to $B_c=4.2+0.5\ln(N)$.
(b)  $dB_c/dN$ for the same data.
Negative values of $dB_c/dN$ indicate clusters that are more stable than
neighboring cluster sizes.
}
\label{fig:stability}
\end{figure}

To measure the stability of the configurations at different values of $N$,
we perform a series of simulations in which $B$ is initially set to a value
which gives a stable configuration, and is then abruptly lowered to a new
value.  We can identify when $B<B_c$ by observing the cluster break into
two or more pieces.  In Fig.~\ref{fig:stability}(a) we plot $B_c$
versus $N$ through $N=80$.  Configurations with $B\ge B_c$ are stable, while
those with $B<B_c$ break apart.  We find that $B_c$ increases approximately
logarithmically with $N$, but that there are noticeable fluctuations
which are particularly pronounced for small $N$.  This is more easily
seen in a plot of $dB_c/dN$, shown in Fig.~\ref{fig:stability}(b).
Magic clusters appear as pronounced dips in $dB_c/dN$.
This general behavior is similar to that observed in artificial atoms,
where a confining potential is present \cite{Classical}.

\begin{figure}
\includegraphics[width=0.5\textwidth]{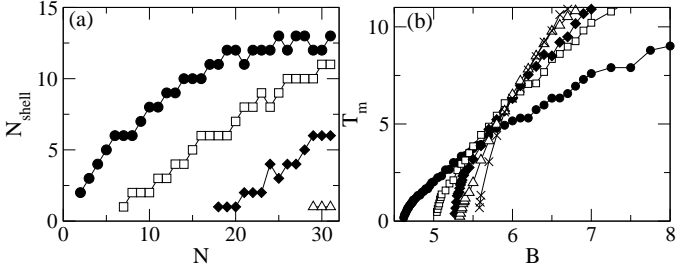}
\caption{(a) The number of colloids $N_{shell}$ in the
outermost shell ($\bullet$),
second shell 
($\square$),
third shell 
($\blacklozenge$),
and fourth shell
($\bigtriangleup$)
as a 
function of $N$.
(b) Disintegration temperature $T_m$ as a function of $B$ for $B>B_c$ and
five different values of $N$. 
$\bullet$: $N=3$.  
$\square$: $N=6$.
$\blacklozenge$: $N=9$.  
$\bigtriangleup$: $N=12$.  
$\times$: $N=15$.
}
\label{fig:Hund}
\end{figure}

The structure of our clusters is predominantly triangular, but we can
define the equivalent of outer and inner shells of colloids in the
clusters.  Unlike systems of particles confined in traps, our shells are
not circular.  The evolution of the shell structure is shown in 
Fig.~\ref{fig:Hund}(a), where the number of colloids $N_{shell}$ in each 
of the first four shells is plotted as a function of cluster size $N$.
The second shell first appears at $N=7$, while the
third shell arises for $N\ge 18$ and the fourth shell for $N\ge 29$.
There is some tendency for the shells to favor an occupancy of $N_{shell}=6$,
as seen by the presence of a step in $N_{shell}$ at $N_{shell}=6$ in
Fig.~\ref{fig:Hund}(a) for all three of the shells.
These results resemble measurements in dusty plasma systems \cite{Juan98}.
We expect the shell structure shown here to be generally applicable to
the type of colloid-colloid interaction we have assumed, and not to be
parameter dependent.

We next consider how to destabilize our cluster structures so that
they fragment.  First we study thermal destabilization.
Here, we fix $B$ at a stable value $B\ge B_c$, and simulate the system
for a fixed number of time steps at each value of $T$,
increasing $T$ until the
thermal fluctuations break the cluster apart.  We perform a series
of runs at each value of $B$ to obtain an average melting temperature
$T_m$ as a function of $B$.  The results for five different cluster
sizes are shown in Fig.~\ref{fig:Hund}(b).  In each case, $T_m$ increases
with $B$ since the attractive energy of the cluster is increasing.  The
rate of increase steepens as $N$ increases.
The destabilization is triggered by the escape of at least one of the
particles from the cluster, and can thus be regarded as a first passage
process that becomes more likely as $T$ increases.  
In our overdamped system, it is difficult for a single particle
to accumulate enough energy to escape the cluster at low temperatures,
unlike the evaporation process that can occur in a massive system.
This tends to stabilize the clusters at low temperatures.

The structures can also be destabilized by starting from a configuration
with $B>B_c$, and then suddenly lowering $B$ to a value $B^*<B_c$.  
This type of process could occur after two nuclei collide and form a
highly compressed but unstable structure, which subsequently fragments.
In a colloidal system, sudden adjustments to the form of the interaction
potential might be achieved through magnetic interactions, as was
considered for purely repulsive particles in Ref.~\cite{Zahn}, or
through electric interactions, as in Ref.~\cite{Yethiraj}.
We begin with configurations prepared slightly above $B_c$ and perform a
series of quenches at small but finite $T$
to different values of $B^*$.  We measure 20 quenches
for each value of $B^*$ in order to accumulate statistics on the resulting
fragment configurations.  The clusters break into smaller clusters which are
stable at $B^*$.  As $B_c-B^*$ increases, the number of 
fragment clusters increases
and the average size of a fragment cluster drops.  

\begin{figure}
\includegraphics[width=0.5\textwidth]{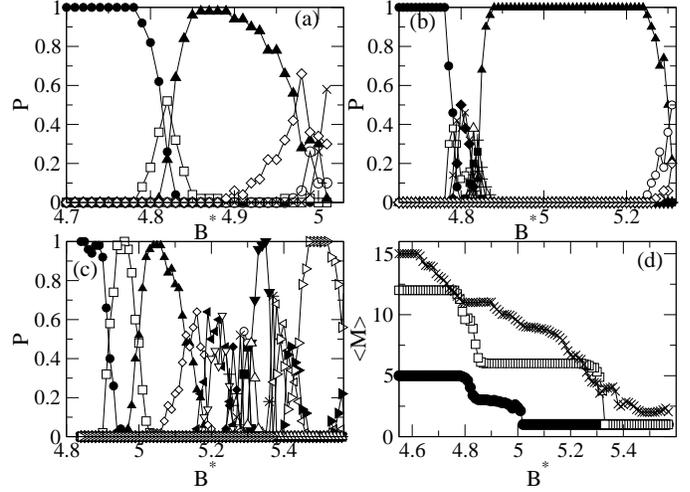}
\caption{(a) Decay modes for $N=5$.  $P$ is the probability of observing
the mode, and $B^*$ is the quench depth.  
$\bullet$: ($1^5$); 
$\square$: (1$^3$,2$^1$); 
$\blacktriangle$: (1$^1$,2$^2$); 
$\diamond$: (2$^1$,3$^1$); 
$\bigcirc$: (1$^2$,3$^1$); 
$\times$: (1$^1$,4$^1$).
(b) Decay modes for $N=12$.  
$\bullet$: (1$^{12}$); 
$\square$: (1$^{11}$,2$^1$); 
$\times$: (1$^{10}$,2$^2$); 
$\blacklozenge$: (1$^6$,2$^3$); 
$\bigtriangleup$: (2$^2$,3$^1$,5$^1$); 
$+$: (1$^4$,2$^1$,3$^2$); 
$\blacksquare$: (1$^5$,2$^2$,3$^1$);
$\blacktriangle$: (1$^3$,3$^3$); 
$\bigcirc$: (1$^2$,3$^2$,4$^1$); 
$\diamond$: (1$^1$,3$^1$,4$^2$).
(c) Major decay modes for $N=15$.  
$\bullet$: (1$^7$,2$^4$); 
$\square$: (1$^6$,2$^3$,3$^1$); 
$\blacktriangle:$ (1$^5$,2$^2$,3$^2$); 
$\diamond$: (1$^4$,2$^1$,3$^3$); 
$\blacktriangleleft$: (1$^3$,2$^1$,3$^2$,4$^1$);
$\bigtriangledown$: (1$^2$,3$^3$,4$^1$); 
$+$: (1$^2$,2$^1$,3$^2$,5$^1$); 
$\blacklozenge$: (1$^1$,3$^3$,5$^1$); 
$\times$: (2$^1$,3$^1$,5$^2$);
$\bigcirc$: (1$^1$,3$^1$,5$^1$,6$^1$); 
$\blacksquare$: (1$^2$,2$^1$,5$^1$,6$^1$);
$\bigtriangleup$: (3$^1$,5$^1$,7$^1$); 
$\blacktriangledown$: (1$^1$,2$^1$,5$^1$,7$^1$); 
$\ast$: (1$^2$,6$^1$,7$^1$); 
$\triangleleft$: (1$^1$,7$^2$); 
$\blacktriangleright$: (2$^1$,6$^1$,7$^1$);
$\triangleright$: (7$^1$,8$^1$).
(d) Average multiplicity $\langle M\rangle$ of the decay as a function of
$B^*$ for $N=5$ 
($\bullet$),
12 ($\square$),
and 15 
($\times$).
}
\label{fig:decay}
\end{figure}

\begin{figure}
\includegraphics[width=0.5\textwidth]{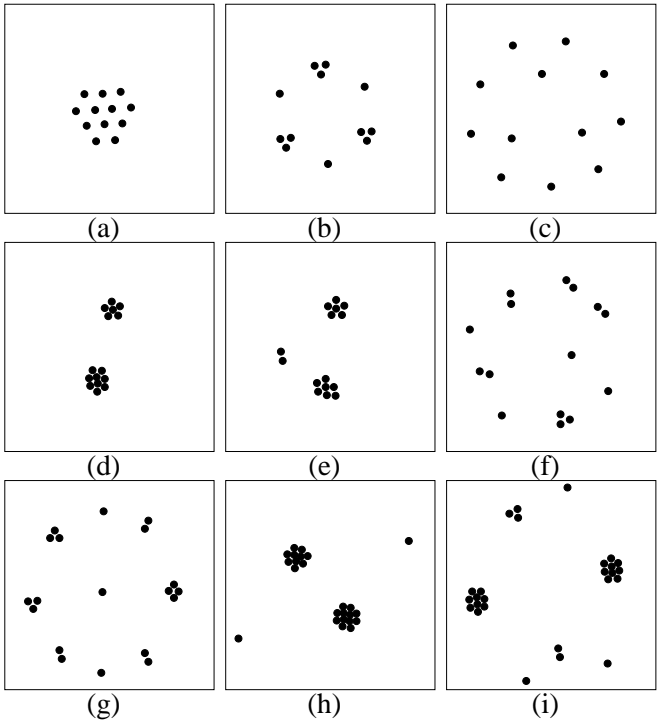}
\caption{Images of decay. 
(a) $N=12$, initial configuration.
(b) $N=12$, $B^*=5$, (1$^3$,3$^3$). 
(c) $N=12$, $B^*=4.75$, (1$^{12}$).
(d) $N=15$, $B^*=5.5$, (6$^1$,9$^1$).  
(e) $N=15$, $B^*=5.35$, (2$^1$,6$^1$,7$^1$).
(f) $N=15$, $B^*=5.1$, (1$^4$,2$^4$,3$^1$).
(g) $N=19$, $B^*=5.55$, (1$^3$,2$^3$,3$^2$,4$^1$).
(h) $N=24$, $B^*=5.7$, (1$^2$,10$^1$,12$^1$).
(i) $N=26$, $B^*=5.6$, (1$^3$,2$^1$,3$^1$,9$^2$).
}
\label{fig:decayimage}
\end{figure}

Different clusters show different types of decay 
modes.  We notate these modes using
$(j_1^{n_{j_1}},j_2^{n_{j_2}},...j_m^{n_{j_m}})$,
where $j_i$ is the number of colloids in clusters of type $i$, 
$n_{j_i}$ is the number of clusters of type $i$, and there are $m$
different cluster types.
We also measure the average multiplicity of the decay at a given
value of $B$, $\langle M(B)\rangle=\langle \sum_{i=1}^{m}n_{j_m}(B)\rangle$.
We illustrate the 
probability $P$ of each decay mode for $N=5$ 
as a function of $B^*$ in Fig.~\ref{fig:decay}(a).
Here we observe all six possible decay modes, and find two dominant
decay modes: 
a breakup into one 1-particle cluster and two 2-particle clusters, denoted
($1^1$,$2^2$), 
or complete disintegration into five 1-particle clusters, denoted
$(1^5)$.  For the magic cluster $N=7$ (not shown),
we find only a single decay mode of complete disintegration, ($1^7$), unless
$B^*$ is extremely close to $B_c$.
The $N=7$ state has such strong symmetry that the particles are not able
to form subclumps as they break away from the main clump.
Fig.~\ref{fig:decay}(b) shows the decay modes 
for the magic cluster $N=12$, illustrated in Fig.~\ref{fig:decayimage}(a).
The two dominant decay modes 
are both symmetric: a breakup into three 1-particle clusters and three 
3-particle clusters ($1^3$,$3^3$), 
illustrated in Fig.~\ref{fig:decayimage}(b), 
and complete disintegration, ($1^{12}$), 
shown in Fig.~\ref{fig:decayimage}(c).  
At $B^*$ values falling in the transition region between these
two regimes for the case of a cluster held at finite temperature, 
numerous decay modes occur with no single mode dominating.  
In the case of an asymmetric, nonmagic cluster such as $N=15$, plotted
in Fig.~\ref{fig:decay}(c), a series of decay modes appears.  Images of
representative decays are shown in Fig.~\ref{fig:decayimage}(d-f).  For
$B^*$ close to $B_c$, the cluster breaks into two large asymmetric pieces,
such as ($6^1$,9$^1$) shown in Fig.~\ref{fig:decayimage}(d).  
As $B^*$ is lowered, smaller pieces
begin to appear, as in the case of (2$^1$,6$^1$,7$^1$)
in Fig.~\ref{fig:decayimage}(e).  For $B^*$
considerably below $B_c$, the cluster disintegrates into many small pieces,
as illustrated in Fig.~\ref{fig:decayimage}(f) for (1$^4$,2$^4$,3$^1$).
Similar patterns occur for larger clusters.  In Fig.~\ref{fig:decayimage}(g),
the magic cluster $N=19$ undergoes a nearly symmetric decay around a single
central particle into the configuration (1$^3$,2$^3$,3$^2$,4$^1$).
A decay into two large pieces combined with two small 
fragments is shown in Fig.~\ref{fig:decayimage}(h) for $N=24$.
The relatively symmetric but elongated $N=26$ cluster is illustrated
breaking into two symmetric large fragments with $N=9$ 
along with a handful of smaller fragments in Fig.~\ref{fig:decayimage}(i).

The average multiplicity of the decay $\langle M\rangle$, 
or the average number of fragments
into which the cluster breaks, increases as $B^*$ decreases.  This is
shown in Fig.~\ref{fig:decay}(d) for $N=5$, 12, and 15.  The presence of
two dominant decay modes for $N=5$ and 12 appears as steps in 
$\langle M\rangle$ as a function of $B^*$.  The asymmetric cluster $N=15$
has a much more complex structure of decay modes, with none of the modes
strongly dominant.  This produces a smoother increase of $\langle M\rangle$
with decreasing $B^*$.  The clusters reach the point of complete disintegration
into individual fragments ($\langle M\rangle=N$) 
at $B^*=4.78$ for $N=5$, $B^*=4.76$ for $N=12$, and $B^*=4.62$ for $N=15$.

In conclusion, motivated by recent experiments in colloidal
systems with competing interactions,
we study a system that can form stable
clusters in the absence of any confining potential.  
We have shown that some cluster sizes
are magic, and correspond to a defect-free segment of a triangular
lattice.  The strength of the short-range attractive term required to
stabilize the clusters increases logarithmically with the cluster
size.  Clusters can be melted by increasing the temperature, or 
destabilized by suddenly quenching the attractive portion of the interaction
to a lower value.  We find a variety of decay modes in the latter case, with
a structure that depends on the symmetry of the original cluster.
These results suggest that colloidal systems 
of self-confined clusters where the interactions can be controlled
may offer insights into the
effects of interactions between charged fragments during decay processes
such as multifragmentation for a larger 
class of systems including molecular clusters
and nuclei. 

\begin{acknowledgement}
We thank W.K. Kegel for useful comments and bringing to our attention some
of the recent experimental work on colloidal systems 
with competing interactions.   
This work was supported by the U.S. Department of Energy
under Contract No. W-7405-ENG-36.
\end{acknowledgement}

\end{document}